\documentclass[twocolumn,aps,showpacs,groupedaddress]{revtex4}
\usepackage{graphicx}
\usepackage{epsfig}
\usepackage{subfigure}

\begin{document}

\title{Feedback control on geometric phase in  dissipative two-level systems}
\author{H. Y. Sun, P. L. Shu, C. Li, X. X. Yi}
\affiliation{School of Physics and Optoelectronic Technology,
              Dalian University of Technology, Dalian 116024, China}

\date{\today}

\begin{abstract}
The effect of feedback on  a two-level dissipative system is studied
in this paper. The results show that it is possible to control the
phase in the open system even if its state can not be manipulated
from an arbitrary initial one to an arbitrary final one. The
dependence of the geometric phase on the control parameters is
calculated and discussed.
\end{abstract}

 \maketitle

Control on a quantum system is a major task required for quantum
information processing. Several approaches to reach this goal have
been proposed in the past decade, which can be divided into open
loop and closed loop problems, according to how the controls enter
the dynamics. In the open loop problems,  the controls enter the
dynamics through the system Hamiltonian. It affects the time
evolution of the system state, but not its spectrum, i.e., the
eigenvalues of the target density matrix $\rho_f$ remain unchanged
in the dynamics  due to the unitarity of the evolution. In the
closed loop problems, a feedback is required from the  controller to
enter the system based on measurement performed  for the open
system. This feedback control strategy can be traced back to
1980's\cite{haus86,yamamoto86,shapiro87} when it is used to explain
the observation\cite{walker85} of subshot-noise fluctuations in an
in-loop photon current. In these works, the authors did this using
quantum Langevin equations and  semiclassical techniques, the latter
approach was made fully quantum-mechanical by Plimak\cite{plimak94}.
For linear systems, all of these approaches and the trajectory
approach\cite{wiseman93,wiseman94,mancini07} are equally easy to use
to find analytical solutions. Nevertheless, the trajectory approach
has advantages that (1) it is applicable for quantum system with
non-linear dynamics, and (2) it is very easy to consider the limit
of Markovian (instantaneous) feedback, i.e.,  a master equation
describing the unconditional system dynamics with feedback may be
possible to derive\cite{wiseman93,wiseman94,mancini07}. The
Markovian feedback can be used to modify the stationary state in a
two-level dissipative quantum system, and this feedback manifests
interesting effects on the time-optimal control in the quantum
system governed by the Lindbald master equation\cite{wang08}.

Geometric phases in quantum theory attracted great interest since
Berry\cite{berry84} showed that the state of a quantum system
acquires a purely geometric feature in addition to the usual
dynamical phase when it is varied slowly and eventually brought back
to its initial form. The Berry phase has been extensively
studied\cite{shapere89,thouless83} and generalized in various
directions\cite{aharonov87,
sjoqvist00l,ericsson03,carollo03,fuentes02}, such as geometric
phases for mixed states \cite{sjoqvist00l}, for open
systems\cite{carollo03}, and with a quantized field
driving\cite{fuentes02}. In a recent paper \cite{sjoqvist00a},
Sj\"oqvist calculated the geometric phase for a pair of entangled
spins in a time-independent uniform magnetic field. This is an
interesting development in holonomic quantum computation and it
shows how the prior entanglements modify the Berry phase. This study
was generalized\cite{tong03} to the case of spin pairs in a rotating
magnetic field, which showed that the geometric phase of the whole
entangled bipartite system can be decomposed into a sum of geometric
phases of the two subsystems, provided the evolution is cyclic. A
renewed interest in geometric phenomena in quantum physics has been
recently motivated by the proposal of using geometric phases for
quantum computing. Geometric phases depend only on the geometry of
the path executed, and are therefore resilient to certain type of
errors. The idea is to explore this inherent robustness provided by
the topological properties of some quantum systems as a means of
constructing built-in fault tolerant quantum logic gates.  Various
strategies have been proposed to reach this goal, some of them
making use of purely geometric
evolution\cite{zanardi99,pachos001,pachos002}. Others make use of
hybrid strategies that combine together geometric and dynamical
evolution\cite{jones99,ekert00}. Several proposals for geometric
quantum computations have been suggested and realized in different
context, including NMR experients\cite{jones99}, ion
traps\cite{duan01,sorensen00,leibfried03,garcia03,staanum04}, cavity
QED\cite{recati02}, atomic ensembles\cite{unanyan99,li04}, Josephson
junction\cite{giuseppe00}, anyonic system\cite{yu03} and quantum
dot\cite{paolo03,carvalho07}.

For open systems governed by the Lindblad master equation, it was
shown that the controls can not fully compensate the effect of
decoherence, indicating that the state of open systems can not be
manipulated from an arbitrary initial state to an arbitrary final
state. This gives rise to a question that can the geometric phase of
such an open system be controlled?

In this paper, we shall study the effect of feedback on the
geometric phase in a dissipative two-level system governed by the
Lindblad master equation. Consider an atom with two relevant levels
$\{|g\rangle, |e\rangle\}$ and lower operator
$\sigma_-=|g\rangle\langle e|.$ Let the atomic decay rate be
$\gamma$ and let it be driven by a classical magnetic field
$\vec{B}(t).$ Within the Markovian approximation for the
system-environment couplings, the time evolution of the two-level
system is described by the Lindblad master equation,
\begin{eqnarray}
i\frac{\partial}{\partial t}\rho&=&\left[H_0,\rho
\right]+\mathcal{L}(\rho),\nonumber\\
H_0&=&\mu\overrightarrow{B}\cdot\overrightarrow{\sigma},\nonumber\\
\mathcal{L}(\rho)&=&i\gamma(F\sigma^-\rho\sigma^+F^\dagger
-\frac{1}{2}\rho\sigma^+\sigma^-
-\frac{1}{2}\sigma^+\sigma^-\rho).\label{me1}
\end{eqnarray}
Here a closed-loop control $F$ is introduced, which is triggered
immediately only after a detection click, namely a quantum jump
occurs. This scheme was used to generate and protect entangled
steady state in cavity QED system and the jump feedback
$F\sigma_-\rho\sigma_+F^{\dagger}$ can be understood as follows. The
unitary operator $F$ is applied only immediately after a detection
event, which is described by  term $\sigma_-\rho\sigma_+$.
Intuitively the stationary states depend on the feedback operator
$F$. So, once the measurement prescription has been chosen, the
freedom to design a feedback to produce a stationary state lies in
the different choices for the feedback operator $F$. Although an
enormous range of possibilities for $F$ is allowed, even considering
the limitations imposed by experimental constraints, we here choose
(with the constraint $FF^{\dagger}$ = 1)
\begin{equation}
F=e^{i\overrightarrow{\sigma}\cdot\overrightarrow{A}}=
\cos{A}+\frac{i\overrightarrow{\sigma}\cdot\overrightarrow{A}}{A}\sin{A},
\label{feedbackF}
\end{equation}
where we denote $A=|\overrightarrow{A}|.$ In fact the feedback $F$
written in this form covers all allowed possibilities. Writing the
reduced density matrix
\begin{equation}
\rho(t)=\frac {1}{2}+\frac
{1}{2}\overrightarrow{p}(t)\cdot\overrightarrow{\sigma}
\end{equation}
we show after a simple algebra that
\begin{equation}
\left[H_0,\rho \right]=\frac{i\mu\overrightarrow{\sigma}}{2}
\cdot\left[\overrightarrow{B}\times\overrightarrow{p}-\overrightarrow{p}\times\overrightarrow{B}
\right],
\end{equation}
and
\begin{eqnarray}
\mathcal{L}(\rho)&=&\overrightarrow{\sigma}\cdot
\{-\frac{i}{2}\gamma\frac{\overrightarrow{p}}{2}
-\frac{i}{4}\gamma\overrightarrow{a}
-\frac{i\gamma}{2}(\frac{1}{2}+\frac{\overrightarrow{p}}{2}\cdot\overrightarrow{a})
\nonumber\\
&\cdot&[\cos{2A}\overrightarrow{a}+\frac{\sin{2A}}{A}\overrightarrow{a}\times\overrightarrow{A}
+2\frac{A_z\sin{A}^2}{A^2}\overrightarrow{A}]\}.\nonumber\\
\end{eqnarray}
Here $\overrightarrow{a}=(0,0,1)$, and the master equation can be
rewritten as,
\begin{widetext}
\begin{equation}
\frac{\partial}{\partial t} \left(
\begin{array}{c}
p_x\\p_y\\p_z
\end{array}
\right) = \left(
\begin{array}{ccc}
-\frac{\gamma}{2} & -2\mu B_z & 2\mu
B_y-\frac{\gamma}{2}[-\frac{\sin{2A}}{A}A_y+2\frac{A_x A_z}{A^2}{\sin^2{A}}] \\
2\mu B_z & -\frac{\gamma}{2} & -2\mu B_x-\frac{\gamma}{2}[\frac{\sin{2A}}{A}A_x+2\frac{A_y A_z}{A^2}\sin^2{A}] \\
-2\mu B_y & 2\mu B_x &
-\frac{\gamma}{2}-\frac{\gamma}{2}[\cos{2A}+2\frac{A_z^2}{A^2}\sin^2{A}]
\end{array}
\right) \left(
\begin{array}{c}
p_x\\p_y\\p_z
\end{array}
\right)
+2\left(\begin{array}{lll}-\frac{\gamma}{4}[-\frac{A_y}{A}\sin{2A}+\frac{2A_xA_z}{A^2}\sin^2{A}]\\
-\frac{\gamma}{4}[\frac{A_x}{A}\sin{2A}+\frac{2A_yA_z}{A^2}\sin^2{A}]\\
-\frac{\gamma}{4}-\frac{\gamma}{4}[\cos{2A}+\frac{2A_z^2}{A^2}\sin^2{A}]
\end{array}\right).
\end{equation}
\end{widetext}
For an open system, its state in general is not pure and the
evolution of the system is not unitary. For non-unitary evolution,
the geometric phase can be calculated as follows. First, solve the
eigenvalue problem for the reduced density matrix $\rho(t)$ and
obtain its eigenvalues $E_k(t)$ as well as the corresponding
eigenvectors $|E_k(t)\rangle$; Second, substitute $E_k(t)$ and
$|E_k(t)\rangle$ into\cite{tong04},
\begin{equation}
\gamma_\textrm{g}(\tau)=\textrm{arg}\sum_k\left[\langle
E_k(t=0)|E_k(\tau)\rangle e^{-\int_0^\tau\langle E_k(t) |
\dot{E}_k(t)\rangle dt}\right].
\end{equation}
where $\gamma_g$ is the geometric phase for the system undergoing
nonunitary evolution\cite{tong04}. In the following we shall choose
\begin{equation}
|\Psi(0)\rangle=\cos{\frac{\theta}{2}}|e\rangle+\sin{\frac{\theta}{2}}|g\rangle
\end{equation}
as the initial state for the dissipative two-level system, in terms
of $\vec{p}=(p_x,p_y,p_z)$ the initial state can be expressed as
\begin{equation}
p_x=\sin{\theta},p_y=0,p_z=\cos{\theta}.
\end{equation}
The final state of the dissipative two-level system is then
\begin{eqnarray}
\rho&=&\left(\begin{array}{cc}\frac{1}{2}+\frac{1}{2}p_z&\frac{1}{2}p_x+\frac{i}{2}p_y
\\ \frac{1}{2}p_x-\frac{i}{2}p_y&\frac{1}{2}-\frac{1}{2}p_z\end{array}\right)\nonumber\\
&=&\frac{1}{2}+\frac{1}{2}\sqrt{p_x^2+p_y^2+p_z^2}\left(\begin{array}{cc}
\cos{\alpha}&\sin{\alpha}e^{i\phi}
\\ \sin{\alpha}e^{-i\phi}&-\cos{\alpha}
\end{array}\right),
\end{eqnarray}
with $\alpha$ and $\phi$ defined by,
\begin{eqnarray}
\cos{\alpha}&\equiv&\frac{p_z}{\sqrt{p_x^2+p_y^2+p_z^2}},\nonumber\\
\tan{\phi}&\equiv&\frac{p_y}{p_x}.
\end{eqnarray}
The eigenvalues and corresponding eigenstates of the reduced density
matrix follows
\begin{equation}
E_{\pm}=\frac{1}{2} \pm\frac{1}{2}\sqrt{p_x^2+p_y^2+p_z^2},
\end{equation}
and
\begin{eqnarray}
|E_+\rangle&=&\cos{\frac{\alpha}{2}}e^{i\phi}|e\rangle
+\sin{\frac{\alpha}{2}}|g\rangle,\nonumber\\
|E_-\rangle&=&\sin{\frac{\alpha}{2}}e^{i\phi}|e\rangle-\cos{\frac{\alpha}{2}}|g\rangle,
\end{eqnarray}
respectively.  It is easy to check that,
\begin{equation}
\frac{1}{2}-\frac{1}{2}\sqrt{p_x^2(0)+p_y^2(0)+p_z^2(0)}=0,
\end{equation}
so the geometric phase $\gamma_g$ reduces to
\begin{equation}
\gamma_\textrm{g}=\textrm{arg}\left[\langle E_+(t=0)|E_+(t)\rangle
e^{-\int_0^\tau\langle E_+(t) | \dot{E}_+(t)\rangle dt}\right],
\end{equation}
straightforward calculations show that
\begin{equation}
\langle E_+(t)|\frac{\partial}{\partial
t}|E_+(t)\rangle=i\cos^2{(\frac{\alpha}{2})}\frac{\partial\phi}{\partial\
t},
\end{equation}
and
\begin{eqnarray}
\frac{\partial\phi}{\partial\ t}
=\frac{\partial\phi}{\partial\cos{\phi}}\cdot\frac{\partial\cos{\phi}}{\partial
t}=-\frac{\sqrt{p_x^2+p_y^2}}{p_y},
\end{eqnarray}
\begin{equation}
\frac{\partial\cos{\phi}}{\partial
t}=\frac{\dot{p_x}\sqrt{p_x^2+p_y^2}-
\frac{p_x(p_x\dot{p_x}+p_y\dot{p_y})}{\sqrt{p_x^2+p_y^2}}}{p_x^2+p_y^2}.
\end{equation}

\begin{figure}
\subfigure{\epsfig{file=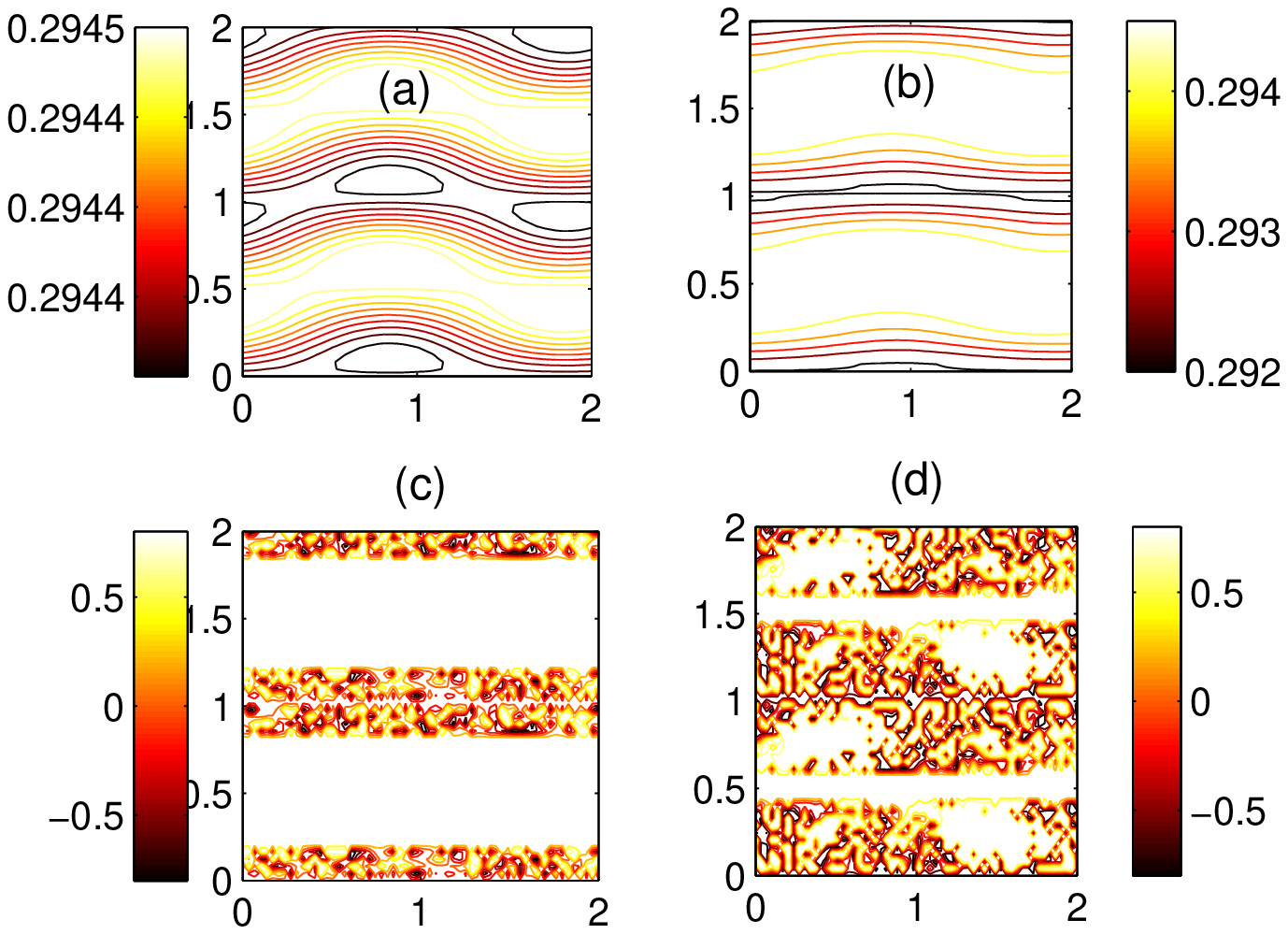,width=0.8\columnwidth,
height=0.6\columnwidth}} \vskip -0.3cm
\subfigure{\epsfig{file=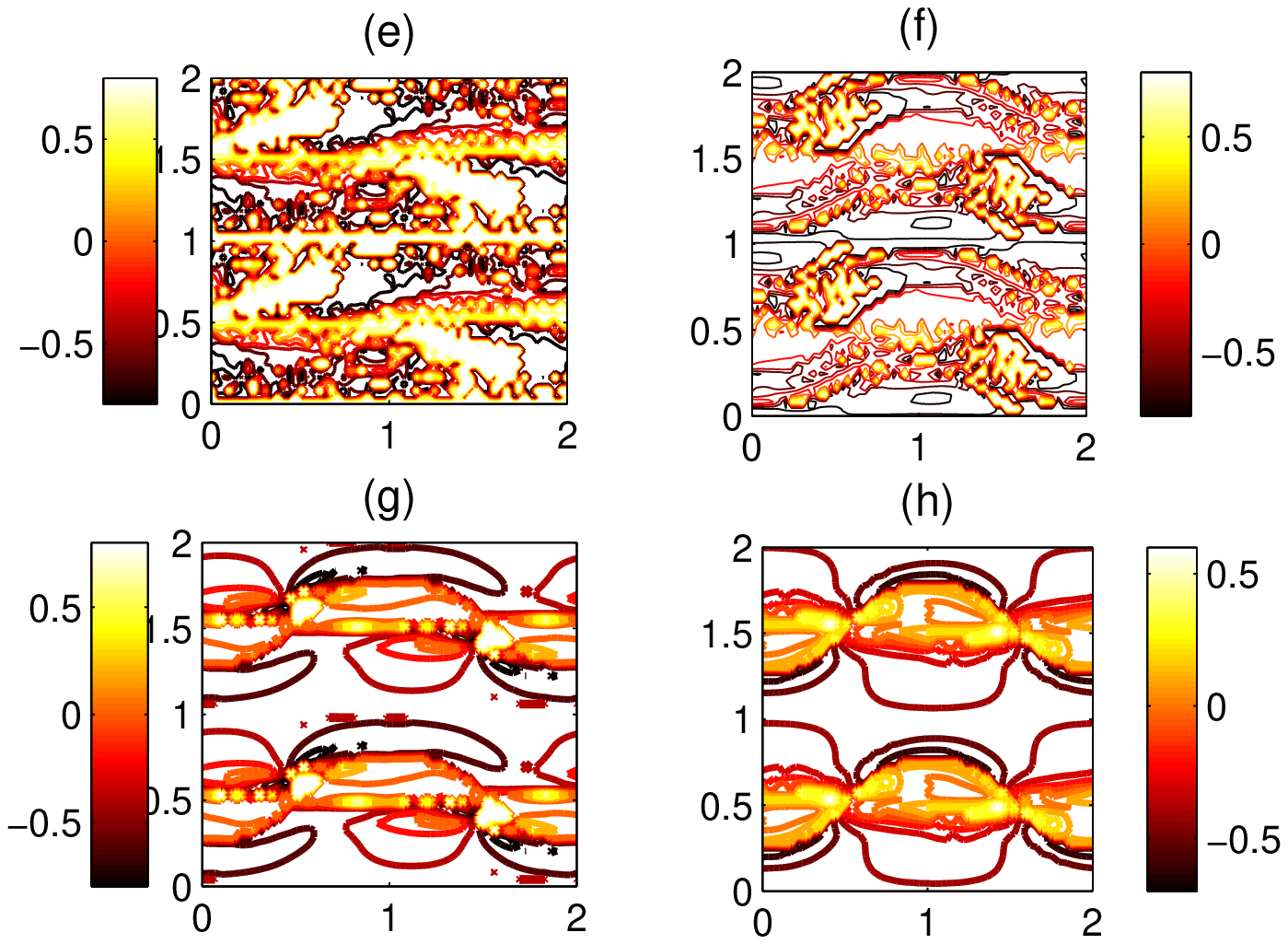,width=0.8\columnwidth,
height=0.6\columnwidth}} \caption{ (color online) Contour plots for
the geometric phase as a function of $A$ (vertical axis) and $\beta$
(horizontal axis). $\omega$ was chosen to be $\omega=0.005\mu B_0,$
and $[0,\tau]$ represents time interval of the system evolution,
$\tau=2\pi/\omega.$ In these plots, we set $\mu B_0=1,$ and both $A$
and $\beta$ are in units of $\pi$. (a) $\gamma=0.001,$ (b)
$\gamma=0.005,$ (c)$\gamma=0.01,$ (d) $\gamma=0.05,$
(e)$\gamma=0.1,$ (f)$\gamma=0.5,$ (g)$\gamma=1$ and  (h)$\gamma=3.$}
\label{fig1}
\end{figure}

\begin{figure}
\includegraphics*[width=0.8\columnwidth,
height=0.6\columnwidth]{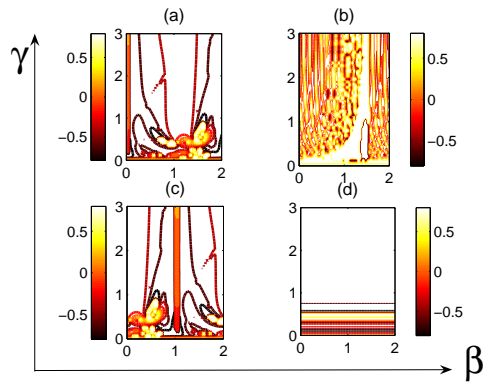} \caption{(color online)
The same as Fig.\ref{fig1}, but with the vertical axis is $\gamma$.
(a)$A=\pi/4$,  (b)$A=\pi/2$, (c)$A=3\pi/4,$ (d)$A=\pi.$  The other
parameters chosen are the same as in Fig.\ref{fig1}} \label{fig2}
\end{figure}
We can use these equations to perform  numerical simulations of the
geometric phase for the open system. In the numerical simulation, we
choose $\vec{B}(t)=B_0(\cos\Theta \cos\Phi, \cos\Theta\sin\Phi,
\sin\Theta)$ as the varying magnetic field with $\Phi=\omega t.$
Without atomic decay, i.e., $\gamma=0$, the two-level system evolves
freely, and its dynamics is governed by $i\partial\rho/\partial
t=[H_0,\rho]$, where $H_0$ is given in Eq.(\ref{me1}). The
instantaneous eigenstates of $H_0$ are
$|+\rangle=\cos{\frac{\Theta}{2}}e^{i\Phi}|e\rangle
+\sin{\frac{\Theta}{2}}|g\rangle,$ $
|-\rangle=\sin{\frac{\Theta}{2}}e^{i\Phi}|e\rangle-\cos{\frac{\Theta}{2}}|g\rangle,$
with the corresponding eigenvalues $e_{\pm}=\pm\mu B_0$,
respectively. For this system to evolve adiabatically, the adiabatic
condition requires $\omega\ll \mu B_0$. In the numerical simulation,
$\omega$ was chosen to be $\omega=0.005\mu B_0,$ and
$\tau=2\pi/\omega.$ To be specific, we set $A_x=A\sin\beta,
A_y=A\cos \beta,$ and $A_z=0,$ where $A$ is a constant. So the
feedback is characterized by $\beta$ and $A.$ The plots presented in
Fig.\ref{fig1} are for the geometric phases acquired by the
dissipative two-level system as a function of $A$ and $\beta$. For
very small atomic decay rate $\gamma\rightarrow 0$
(Fig.\ref{fig1}-(a)), the geometric phase approaches  a constant
$\gamma_g\sim (1-\cos\theta)$ (in units of $\pi$). As $\gamma$
increases, the range of the geometric phase acquired by the
dissipative
 system increases, implying that the geometric phase can be
 controlled even with large atomic decay rate $\gamma$. This is different
 from the control on quantum states, where the control can not fully
 compensate the decoherence. Figure \ref{fig1} also shows that the
 geometric phase is a periodic function of $A$, this can be understood by
 examining  Eq.(\ref{feedbackF}), where the feedback control is given.
 In addition to the
 above observation, we can find from figure \ref{fig1} that the geometric
 phase is regular for small and large $\gamma$, while it is irregular for
 intermediate values of $\gamma$. The physics behind this feature is the following.
 For very small $\gamma$, $\mathcal{L}(\rho)$ is negligible, hence
 $H_0$  dominates over $\mathcal{L}(\rho)$
 in the dynamics and the geometric phase is mainly
 determined by $H_0$. When $\gamma$ is large enough such that $\mathcal{L}(\rho)$
 dominates the dynamics, the geometric phase then comes from the
 dynamics governed by $\mathcal{L}(\rho).$
With a specific $A=\pi/4$, the geometric phase as a function of
$\beta$ and $\gamma$ is plotted in figure \ref{fig2}. In all these
plots, we set $\Theta=\theta$, thus the initial state is an
eigenstate  of the free Hamiltonian $H_0$. For $A=\pi/2$, the
feedback is $F=i(\sin\beta\sigma_x+\cos\beta\sigma_y)$ whereas for
$A=\pi$, $F$ becomes 1, i.e., there is no feedback operating on the
system. This can be found in Fig.\ref{fig2}-(d), where the geometric
phase acquired is independent of $\beta.$ From figure \ref{fig2}-(a)
and \ref{fig2}-(c) we find that figure \ref{fig2}-(c) is exactly the
same as figure \ref{fig2}-(a) by replacing $\beta$ by $\beta+\pi$,
indicating that the geometric phase remains unchanged with
$A\rightarrow \pi-A$ and $\beta \rightarrow \beta+\pi.$ This feature
can be understood  as follows. Recall that $\vec{A}=(A\sin\beta,
A\cos\beta,0),$ $F$ can be written as $F=\cos
A+i(\sigma_x\sin\beta+\sigma_y\cos\beta)\sin A,$ it is clear that
$F$ remains unchanged by replacing $A$ and $\beta$ with $\pi-A$ and
$\pi+\beta$, respectively.

To sum up, in this paper, we have studied the effect of feedback on
the geometric phase of a dissipative two-level system. The
dependence of the phase on the feedback parameters are calculated
and discussed. The results suggested that we can manipulated the
phase by a properly designed feedback control. For small and large
atomic dissipative rates with respect to the amplitude of the
driving magnetic field $\mu B_0$, the geometric phase is a periodic
function of the feedback parameters, the physics behind these
features is also presented.

This work was supported by   NSF of China under Grant  No. 10775023.\\

\end{document}